# Influence of CaO and SiO$_2$ additives on the sintering behavior of Cr,Ca:YAG ceramics prepared by solid-state reaction sintering.

M.A. Chaika[a,b], G. Mancardi[c], O.M. Vovk[b]

[a]*Institute of Low Temperature and Structure Research Polish Academy of Science, Okolna 2, 50-422 Wroclaw, Poland.*

[b]*Institute for Single Crystals, National Academy of Sciences of Ukraine, 60, Nauky Avenue, Kharkiv, 61072, Ukraine.*

[c]*Politecnico di Torino, Corso Duca Degli Abruzzi 24, 10129, Torino*

**Abstract**

Optical ceramics of yttrium aluminum garnet (Y$_{3-x}$Ca$_x$Al$_{4.95}$Cr$_{0.05}$O$_{12}$) doped with 0.1 at.% of chromium ions and 0.5, 0.8 and 1.2 at.% of calcium ions were made by reactive sintering in vacuum at 1750°C. The influence of SiO$_2$ and CaO sintering aids on optical properties and microstructure of Cr,Ca:YAG ceramics was investigated. A decrease in CaO concentration led to a reduction of the Cr,Ca:YAG ceramics transparency from 81 to 0 % at 1064 nm, moreover, ceramics with a lower concentration of CaO have abnormal grains and high-level porosity. The obtained features were attributed to the interactions between silica and calcium oxide during vacuum sintering of Cr,Ca:YAG ceramics. We propose and discussed two mechanisms of CaO, SiO$_2$ influence on the formation of Ca,Cr:YAG ceramics. The first one involves the appearance of a liquid phase due to the CaO-SiO$_2$ interaction at a specific ratio under the heating stage of ceramics sintering. Under this way, the abnormal growth of ceramics grains occurs. The second mechanism consists on the mutual consumption of Si$^{4+}$ and Ca$^{2+}$ ions under the isothermal stage of ceramic sintering which decreases the Cr$^{4+}$ concentration. Both mechanisms have a negative effect on the optical and laser properties of the ceramics.

1. **Introduction**

During the last few decades, YAG based solid-state materials doped with $Cr^{4+}$ have been widely used and investigated as tunable solid-state lasers for the 1.35-1.55 μm spectral region [1] and as passive Q-switch lasers [2, 3, 4]. Much attention has been given to the $Cr^{4+}$:YAG saturated absorber used in microchip Q-switched lasers, which allows generating ultra-short (up to few dozen picoseconds) high power in the maximum pulse. YAG ceramics have not yet been extensively employed for microchips, and, among the advantages, they withstand high concentration of doping ions, the fabrication is easy and they are high-temperature resistant.

The main characteristic of $Cr^{4+}$:YAG materials is the necessity to use divalent additives for charge compensation. Yttrium-aluminum-garnet crystal (YAG) has a cubic structure and belongs to the Ia-3d space group with the stoichiometric formula $[C]_3[A]_2[D]_3O_{12}$, where C, A and D denote dodecahedral, octahedral, and tetrahedral lattice sites, respectively [5, 6]. The possibility of using $Cr^{4+}$:YAG materials as Q-switched lasers is based on the large ground-state absorption cross-section and excited-state lifetime of tetrahedral coordinated $Cr^{4+}$ ions. During vacuum sintering, Cr dopants incorporate into the YAG lattice in the trivalent state, replacing $Al^{3+}$ in the octahedral positions [7, 8]. After vacuum sintering, only $Cr^{3+}$ exists, because the formation of $Cr^{4+}$ requires high-temperature air annealing. $Cr^{4+}$:YAG ceramics must contain divalent additives such as $Ca^{2+}$ or $Mg^{2+}$ for stabilizing $Cr^{4+}$ in the tetravalent state [7, 9].

One of the ways to create highly transparent YAG ceramics is using $SiO_2$ sintering aids. Ikesue et al. [10] first reported highly efficient laser gain of transparent Nd:YAG ceramics produced by sintering a powdered mixture of $Al_2O_3$, $Y_2O_3$, and $Nd_2O_3$ at 1750°C for 8 h. They found that the addition of 0.14 wt.% (1.35 mol%) $SiO_2$ is crucial for sintering transparent Nd:YAG ceramics. This phenomenon has been later confirmed and, nowadays, different concentrations of $SiO_2$ are commonly added as a sintering aid to produce Nd:YAG ceramics of laser-quality [11-14]. The most popular model for describing the interaction between $SiO_2$ and YAG is the formation of a liquid phase [15] or an increase in the concentration of cation vacancies [16] due to the incorporation of $Si^{4+}$ ions into the YAG lattice.

The presence of $SiO_2$ has a negative effect on the properties of $Cr^{4+}$:YAG ceramics, as it leads to a decrease in optical transparency [17] and inhibits the $Cr^{3+} \rightarrow Cr^{4+}$ ions conversion efficiency [18], consequently, it is not convenient to produce $Cr^{4+}$:YAG ceramics. The reason for such behaviour has not yet been revealed. Understanding the interactions between $SiO_2$ and CaO during vacuum sintering will allow expanding our knowledge about ceramics sintering. Moreover, shedding light on these interactions enables to understand how to use CaO and $SiO_2$ additives to create doped $Cr^{4+}$:YAG ceramics for laser applications.

Therefore, the purpose of this work is to investigate the interaction between $SiO_2$ and CaO compounds during vacuum sintering of Ca,Cr:YAG ceramics.

## 2. Experimental

In the present work, high purity $Al_2O_3$ (purity >99.99%, Baikowski, d=0.15-0.3 $\mu$m), $Y_2O_3$ (purity >99.999%, Alfa Aesar, d=<10 $\mu$m), $Y_2O_3$ (purity >99.97%, Alfa Aesar, d=<100 nm), CaO (purity >99.999%, Sigma Aldrich, d=<0.1 $\mu$m) powders were used as starting materials. Tetraethoxysilane (TEOS, purity >99.999%, Sigma-Aldrich, USA) was used as an original substance of $SiO_2$ for sintering aid. Powders were taken in the YAG stoichiometric ratio, the concentration of $Ca^{2+}$ and $Cr^{3+}$ was calculated to replace $Y^{3+}$ and $Al^{3+}$ respectively.

The powder mixtures were ball milled in anhydrous alcohol for 15 hours using high purity $Al_2O_3$ balls, then dried at 70°C and granulated by sieving. The compacts were prepared by uniaxial pressing at P = 250 MPa. Before vacuum sintering, they were annealed in the air for two hours at 600°C. Sintering was performed in a vacuum furnace at 1750°C and a vacuum of $10^{-5}$ Pa for 10 hours. For recharging $Cr^{3+}$ in its tetravalent state, air annealing at 1490°C for 20 hours was performed. The samples of Cr,Ca:YAG ceramics were obtained with a diameter of 10 mm and a thickness of 1.2 mm.

Cr,Ca:YAG ceramics doped with different concentration of calcium 0.5, 0.8, 1.2 at% were labelled as Ca=0.5, Ca=0.8 and Ca=1.2 respectively. Fig. 1 reports a photograph of the mirror-polished Cr,Ca:YAG ceramics after vacuum sintering and followed air annealing. The concentration of chromium oxide was constant and set at 0.1 at.%. All samples were doped with 0.5 wt.% TEOS which yields silica levels of approximately 0.14 wt.%.

The linear optical transmission spectra of Cr,Ca:YAG ceramics were collected with the UV/Vis spectrophotometer "Analytik Jena specord 200" at the range 190-1100 nm. Cr,Ca:YAG ceramics morphology was studied using a JEOL JSM- 6390LV scanning electron microscope, the elemental analysis was performed on Silicon Drift Detector (SSD) X-MaxN Oxford Instruments. To obtain XRD spectra, a DRON 3 type diffractometer was used. The measurement of the variation in elements concentration was carried out by chemical analysis under each stage of ceramics production; the methodology is described in detail in ref. [19]

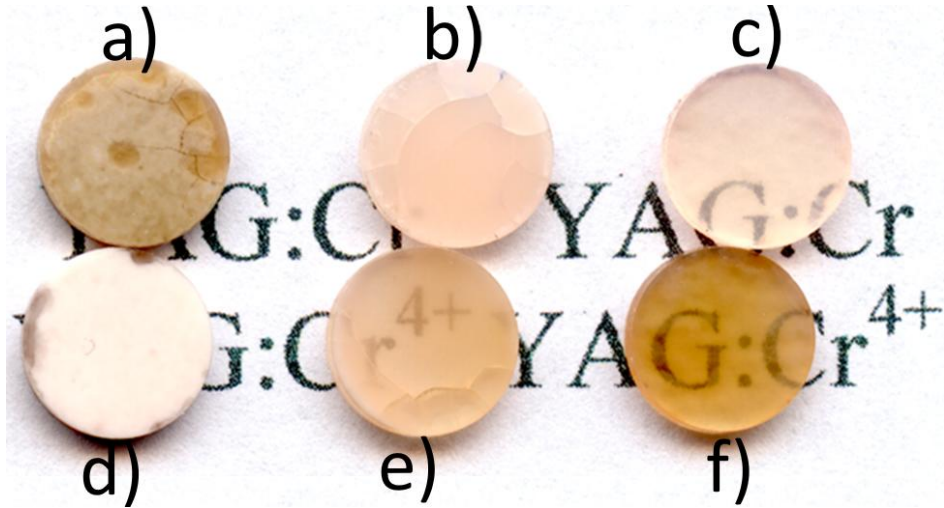

Figure 1: Photograph of the mirror polished a) Ca=0.5, b) Ca=0.8 and c) Ca=1.2 samples after vacuum sintering and d) Ca=0.5, e) Ca=0.8 and f) Ca=1.2 after followed air annealing at 1490°C for 20 h.

## 3. Results

The aim of this work is to understand the influence of dopant interactions on YAG ceramics prepared by solid-state reaction (SSR) sintering in a vacuum, taking as a reference the interaction between CaO and $SiO_2$. The vacuum sintering process can be separated into two stages: heating and isothermal annealing. During the heating stage, the ceramic density increased from 70 to 99.9 %, while only a fraction of the porosity was removed during isothermal annealing [16]. Therefore, the heating stage is a pivotal step in ceramics sintering, and we speculate that the CaO - $SiO_2$ interaction at an early preparation stage can significantly change the final microstructure of the YAG ceramics obtained after sintering. Below, the properties of Cr,Ca:YAG ceramics with different $CaO/SiO_2$ ratio are presented, we expect that the dissimilar properties exhibited by the samples are ascribable to a different interaction between CaO and $SiO_2$ during the initial stage.

*3.1 Structural characterization*

The X-ray diffraction patterns of the prepared samples coincided with the crystal structure of pure

YAG phase and corresponded to the Inorganic Crystal Structure Database (ICSD) Card N 170158 (Fig. 2). New peaks and/or peak shifts were not observed in the Cr,Ca:YAG diffraction pattern compared to the pure YAG one. It can be noted that the sensitivity of XRD analysis was close to 1 at.%, indicating that a small concentration of impurities might be present in the samples.

Our experiment shows that increasing the concentration of Ca from 0.5 to 0.8 led to a stabilization of the grain size and inhibited anomalous grain growth. The grain structure of Cr,Ca:YAG ceramics was investigated by SEM analysis after thermal etching at 1490°C for 20 hours in air ambient. Ca=0.5 sample had a bimodal grain size distribution: most of the grains were in the range 0.5-4.5 µm, with peaks also at 1 µm and 2 µm, whereas larger grains from few dozens to few hundred microns were also detected. An increase in Ca concentration led to a reduction in grain size and to the disappearance of the abnormally large grains. Ca=0.8 and Ca=1.2 samples have grain size distribution from 0.5 to 2 µm with average grain size 0.9 ± 0.05 µm (Fig. 3a). The standard deviation was calculated using the Student's t-test.

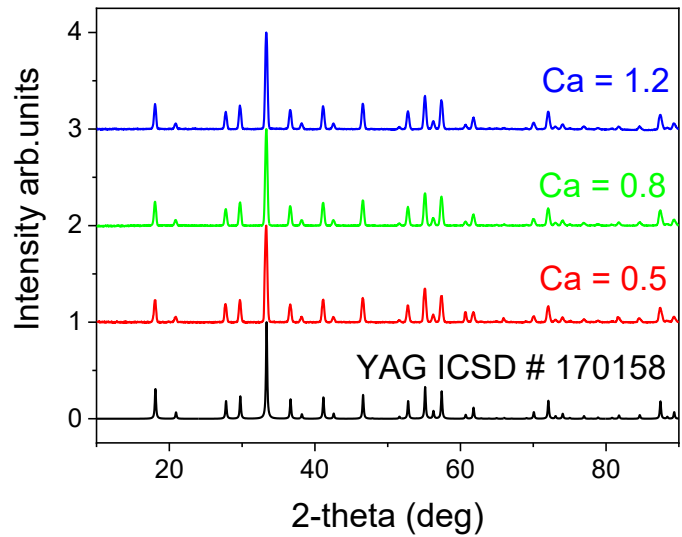

Figure 2: The X-ray diffraction patterns of the prepared samples and pure YAG ICSD reference pattern (Card N 170158).

The SEM analysis shows the appearance of a precipitate on the surface of Ca=0.8 and Ca=1.2 samples after air annealing. It means that the microstructure of the samples correlates with the presence of additives on the surface. Ca=0.5 sample has a clean surface (Fig. 1b), while in the case of Ca=0.8 sample, we observed the formation of a precipitate covering 0.5% of the surface and organized in round shapes with size 1-2 μm (Fig. 1c). Rising the concentration of Ca to 1.2 led to a change in the shape of the precipitate to dendrites and increasing its size up to 70 μm with 7% covering of the surface (Fig. 1d).

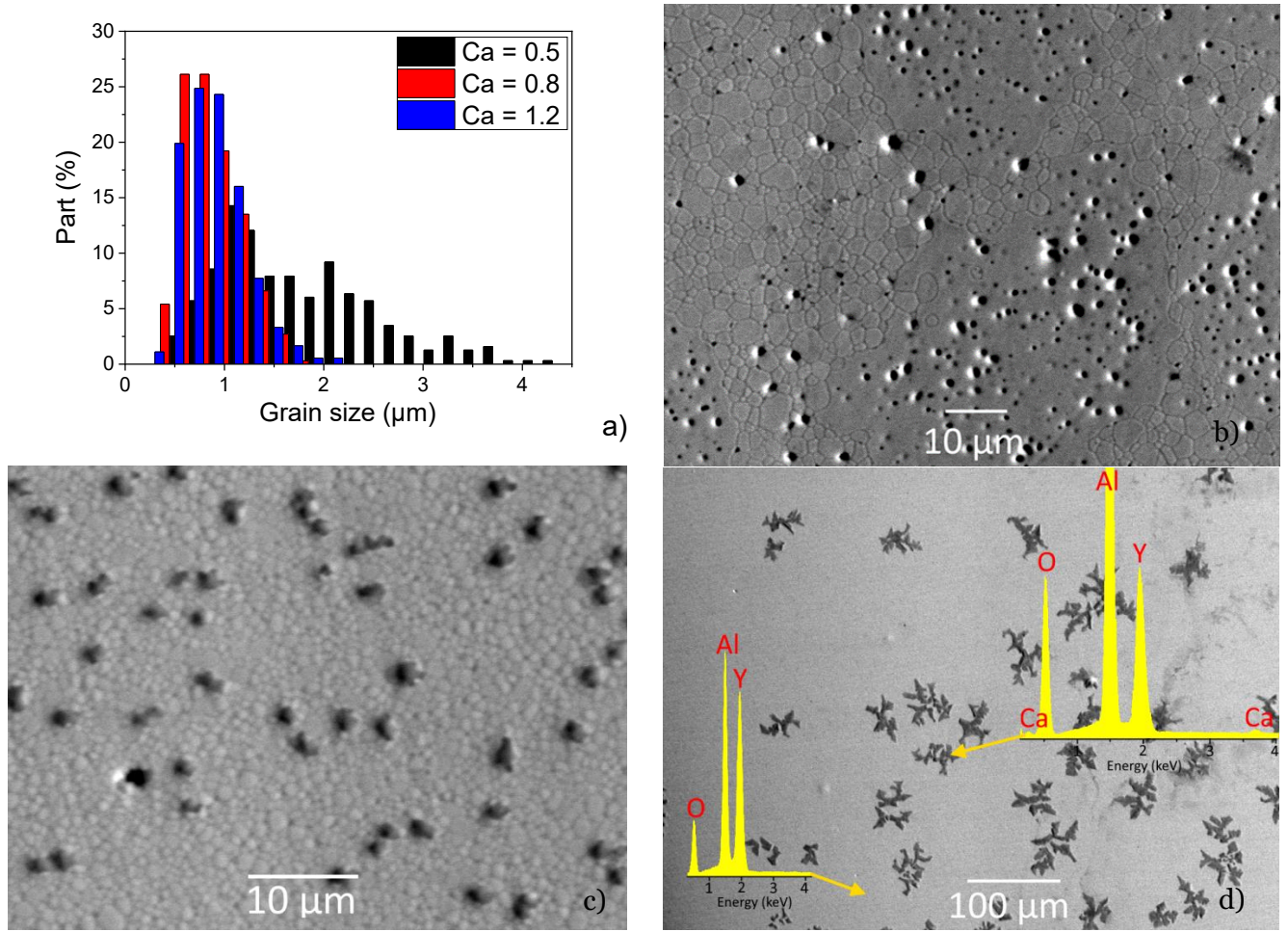

Figure 3: a) Grain size distribution of the samples. c) SEM image surface of Ca=0.5 sample after air annealing at 1490°C for 20 hours. b) SEM image surface of Ca=0.8 sample after air annealing at 1490°C for 20 hours. d) SEM image surface of Ca=1.2 sample after air annealing at 1490°C for 20 hours

We suppose that the observed precipitate originates from the bulk of the ceramics and has grown as islands of thin films on the sample surface. These observed impurity phases in the SEM images look darker than the YAG matrix background, indicating the presence of a high concentration of ions lighter than $Y^{3+}$ or $Al^{3+}$ in the precipitate.

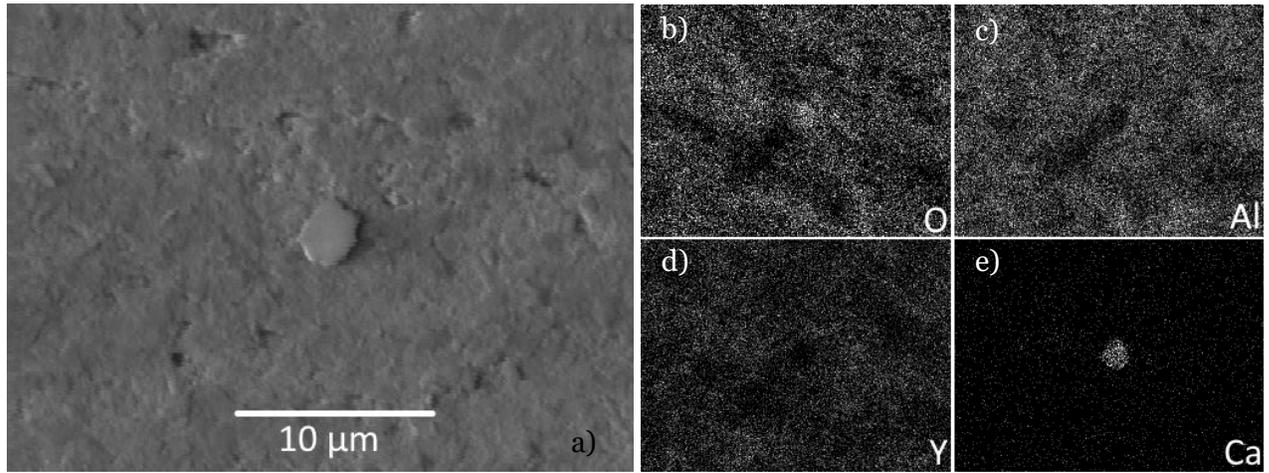

Figure 4: Precipitate on the surface of ceramics of Cr,Ca:YAG (1.2 at% Ca) after air annealing at 1490°C for 20 hours:(a) SEM surface image, EDS mapping of: (b) O, (c) Al, (d) Y, (e) Ca.

Fig. 4 shows the maps of the area with precipitate particles on the surface of Ca = 1.2 sample by EDS analysis of selected elements: Al, Ca, O, and Y. Only basic elements of YAG: Al, O and Y were detected in the area of pure surface, whereas Ca was revealed in the precipitate. The EDS analysis depth is 1-3 µm, and the Y and Al signals coming from the YAG matrix penetrate through the precipitated particle lying on the surface. Evidently, these particles are of submicron thickness and contain Ca.

Summarizing, we detected the appearance of a precipitate on the samples' surfaces. After vacuum sintering, the surface of the samples was clean without any precipitates. Annealing in air leads to the appearance of the precipitate on the surface of Ca=0.8 and Ca=1.2 samples. We suppose that the diffusion of $Ca^{2+}$ ions from the bulk to the ceramic surface during air annealing leads to the formation of a Ca-rich phase. After softly polishing of samples, the surface became clean again, meaning that the Ca-phase is extremely thin.

Probably, the presence of a Ca-rich phase at the grain boundary in Ca=0.8 and Ca=1.2 samples causes the suppression of grain growth in these samples compared to the low calcium content Ca=0.5 sample.

*3.2 Chemical analysis*

The YAG ceramics sintering takes place in several steps and occurs at extremely high temperatures for a long time, meaning that the chemical composition of the samples can change during the process. The chromium and calcium compounds have the highest saturated vapor pressure among all components. Fig. 5 shows a dependence of the initial compounds of the Cr,Ca:YAG ceramics powder mixture saturated vapor pressure on the temperature [20]. It is worth to note that chromium oxide and calcium oxide possess a vapor pressure few orders in magnitude higher than yttrium and aluminum oxides.

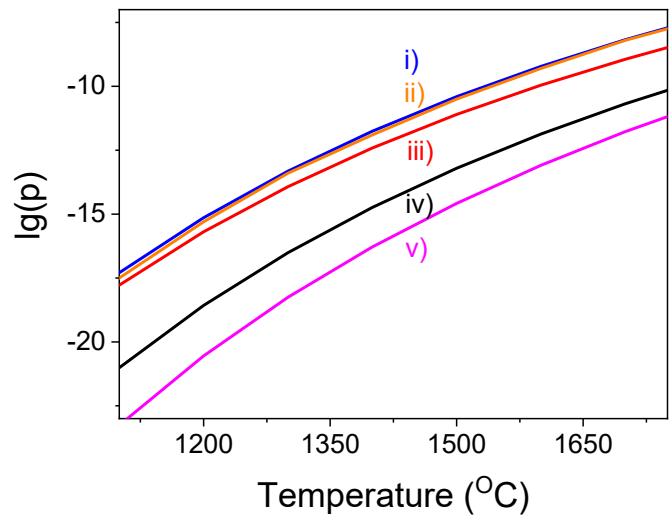

Figure 5: Temperature dependence of saturated vapor pressure of initial compounds: i) $Cr_2O_3$ ii) $SiO_2$ iii) CaO iv) $Al_2O_3$ v) $Y_2O_3$ [20]

A variation of chromium and calcium content in the Cr,Ca:YAG ceramics due to evaporation under vacuum synthesis can strongly affect the sintering process and the functional properties of Cr,Ca:YAG ceramics. The Ca=0.8 sample was investigated by means of elemental chemical analysis. In particular, we analyzed the change in the original powders content after thermal treatment of the initial powder mixture at 600°C, followed by vacuum sintering and air annealing. Table 1 reports the chemical analysis of Ca=0.8 sample, where the steps are indicated as follow: "Powder" - powders mixture, "600°C" - a mixture of the original powder thermally treated at 600°C, "1750°C" – Cr,Ca:YAG ceramics after vacuum sintering.

Table 1: The chemical analysis of Ca=0.8 sample.

| Stage | Element content, wt.% | | | |
|---|---|---|---|---|
| | Y | Al | Ca | Cr |
| Powder | 43.6 | 22.1 | 0.18 | 0.038 |
| 600°C | 44.5 | 22.7 | 0.17 | 0.038 |
| 1750°C | 39.8 | 21.1 | 0.14 | 0.035 |

We detected a decrease in Ca and Cr amounts in the sample during vacuum sintering. The initial element concentrations after thermal treatment at 600°C did not practically change compared to the powder mixture according to the results of the chemical analysis. But the vacuum sintering process led to a decrease in Cr and Ca content of 14 % and 7 % respectively.

### 3.3 Concentration of $Cr^{3+}$ and $Cr^{4+}$ ions

We found that vacuum sintering induces the formation of $Cr^{4+}$ ions. Transmission spectra of the samples after vacuum sintering show only the formation of $Cr^{3+}$ ions (see fig. 6). The two broad absorption bands correspond to the $^4A_2 \rightarrow {}^4T_2$ and $^4A_2 \rightarrow {}^4T_1$ transitions centred at 590 and 430 nm respectively, which are responsible for the bright green color of $Cr^{3+}$:YAG ceramics. After vacuum sintering, Ca=1.2, Ca=0.8 and Ca=0.5 samples had transmittance of 81, 24 and 0 % respectively at $\lambda$=1064 nm.

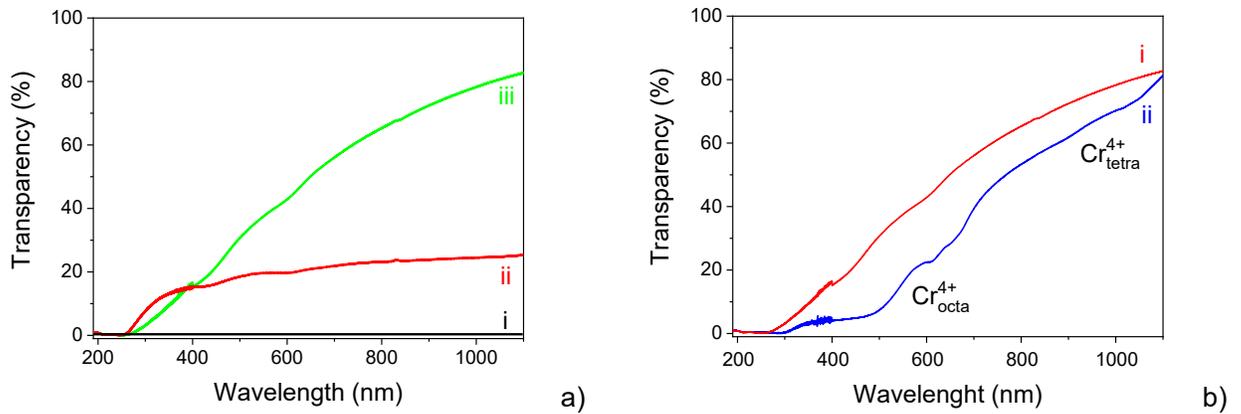

Figure 6: a) Optical in-line transmittance spectra of i) Ca=0.5, ii) Ca=0.8 and iii) Ca=1.2 samples, b) comparison of transmittance spectra between a fully reduced and a fully oxidized Ca=1.2 sample.

Cr,Ca:YAG ceramics transparency decreased after annealing in air at 1490°C for 20 hours due to the formation of $Cr^{4+}$ absorption centres in visible and near-infrared spectral region. The formation of $Cr^{4+}$ ions in Cr,Ca:YAG ceramics takes place during annealing in an oxidizing ambient where the tetravalent additives can be stabilized through the formation of structural defects in the lattice such as divalent additives [21,22].

The concentration of $Cr^{3+}$ and $Cr^{4+}$ ions in different lattice positions for Cr,Ca:YAG ceramics was calculated from the $Cr^{4+}$ absorptions spectra, for a detailed explanation of the methodology, refer to [23]. The calculated $Cr^{4+}$ concentrations are reported in Table 2. According to the calculations, Ca=1.2 sample contains $Cr_A^{4+}$ and $Cr_D^{4+}$ ions close to $9.8 \cdot 10^{17}$ cm$^{-3}$ and $3.6 \cdot 10^{17}$ cm$^{-3}$ respectively. In the case of sample Ca=0.8, the amounts of $Cr_A^{4+}$ and $Cr_D^{4+}$ ions were $3.8 \cdot 10^{17}$ cm$^{-3}$ and $1.5 \cdot 10^{17}$ cm$^{-3}$, respectively. The $Cr^{3+}$ ions concentration was corrected for the $Cr_2O_3$ losses determined by elemental chemical analysis and the amount of $Cr^{4+}$ ions. The sample Ca=1.2 has a higher concentration of $Cr^{4+}$ ions both in the octahedral and in the tetrahedral sites in comparison to the other samples. The $Cr^{4+}/Cr_{total}$ ratio is linearly dependent on $Ca^{2+}$ concentration and we found it to be 6.6 %, 2.6 % and 0 % for samples Ca=1.2, Ca=0.8 and Ca= 0.5. Less than 1% of $Ca^{2+}$ ions are involved in the formation of $Cr^{4+}$ ions, which is much less than what found for samples without $SiO_2$ [23].

Table 2: Parameters of the samples: assignment, $N_0$ - concentration of absorbing center, $N_D/N_A$ - tetrahedral to octahedral $Cr^{4+}$ ions concentration ratio.

| Sample | $N_0 \cdot 10^{17}$, cm$^{-3}$ | | | part of $Cr^{4+}$ ions, % | | $Cr^{4+}$ ratio |
|---|---|---|---|---|---|---|
| | $Cr_A^{3+}$ | $Cr_A^{4+}$ | $Cr_A^{4+}$ | $Cr^{4+}/Ca^{2+}$ | $Cr^{4+}/Cr_{total}$ | $N_D/N_A$ |
| Ca = 1.2 | 190 | 9.8 | 3.6 | 0.8 | 6.6 | 0.37 |
| Ca = 0.8 | 199 | 3.8 | 1.5 | 0.5 | 2.6 | 0.39 |

## 4. Discussion

In this paper, we examined the influence of dopants and their interaction on optical properties and microstructure of YAG-doped ceramics. We suppose that one of the reasons for the dramatic change in microstructure and transparency of the samples is the influence of CaO and $SiO_2$ interactions on the stability of the pore network. Solid-state vacuum sintering of Cr,Ca:YAG ceramics can be considered as a two-stage process: heating and isothermal annealing. During the first stage, the following modifications take place: (i) formation of necks between grains, (ii) necks evolution into grain boundaries, (iii) formation of an open pore network, (iv) thinning of the diameter of the pore channels, (v) their collapse into closed pores [24]. The critical diameter of the pore channels depends on the relation between grain boundary mass transport and surface mass transport before collapsing into close porosity. Smaller pores are removed more easily from the ceramic bulk than larger pores. Therefore, the formation of different inter-grains phases (*i.e.* in the case of CaO and SiO$_2$ interactions) can change the diameter of the pore channels, thus affecting the properties of the final Cr,Ca:YAG ceramics [25]. Our Discussion section is divided into two subsections: in the first one (4.1), the influence of CaO and SiO$_2$ interactions during the heating stage of sintering is presented; in the second one (4.2), the influence of Ca$^{2+}$ and Si$^{4+}$ ions interaction on sample properties is reported.

*4.1 Influence of CaO and SiO$_2$ interactions*

The interactions between CaO and SiO$_2$ dopants led to the formation of several intermediate phases characterized by different Ca-doping levels, the presence of these phases explains the observed properties of the samples. During vacuum sintering, the initial compounds (Al$_2$O$_3$, Y$_2$O$_3$) and dopants (Cr$_2$O$_3$, CaO, SiO$_2$) interact, leading to the formation of intermediate phases which can alter the sintering behavior. The amount of CaO and Cr$_2$O$_3$ was negligibly small compared to Al$_2$O$_3$ and Y$_2$O$_3$ and the amount of CaO was much higher than Cr$_2$O$_3$ while the concentrations of CaO and SiO$_2$ are comparable. Moreover, the CaO and SiO$_2$ low solubility in YAG is in contrast with Cr$_2$O$_3$, which is well soluble in YAG. Therefore, a variation in CaO concentration did not have any influence on its interaction with Al$_2$O$_3$, Cr$_2$O$_3$ and Y$_2$O$_3$ but, in the case of CaO-SiO$_2$ interactions, the opposite behavior is observed.

We suppose that the properties of the samples can be explained by the different interactions between CaO and SiO$_2$ taking place at the YAG grain boundaries during the heating stage. Due to the low solubility of Ca$^{2+}$ and Si$^{4+}$ ions, only a fraction of the CaO and SiO$_2$ can dissolve into the YAG matrix. CaO and SiO$_2$ dopants are located at the grain boundaries during the heating stage of vacuum sintering and the interaction between these dopants leads to the formation of different intermediate compounds, which chemical composition can be predicted from the SiO$_2$-CaO phase diagram by considering the initial SiO$_2$/CaO ratio. It should be noted that correlating the CaO concentration with the optical properties of the samples based only on the binary phase diagram of CaO-SiO$_2$ is not correct: the entire CaO-Cr$_2$O$_3$-Al$_2$O$_3$-Y$_2$O$_3$-SiO$_2$ system should be considered, which is a too complicated system. Although the binary phase diagram of CaO-SiO$_2$ does not provide a true representation of the system, it does allow estimating its behavior.

The SiO$_2$-CaO ratio in Ca=0.5, Ca=0.8 and Ca=1.2 samples were 37-63 mol. %, 26-74 mol. % and mol. 20-80%. According to the CaO-SiO$_2$ phase diagram, [26] the interaction between CaO and SiO$_2$ in the Ca=0.5 sample can lead to the formation of a mixture of rankinite (Ca$_3$Si$_2$O$_7$) and wollastonite (CaSiO$_3$), which eutectic composition melts at 1450 °C. We suppose that the formation of a liquid near 1450°C leads to an early collapsing of the pore channel, which results in the formation of large pores, detected by SEM (see fig. 3b). In the case of Ca=0.8 and Ca=1.2 samples, the lowest temperature at which the liquid can appear is 2057 °C, when the eutectic composition of larnite (Ca$_2$SiO$_4$) and CaO is much higher than the ceramics sintering temperature (1750°C). We assume that the absence of a liquid phase is the reason for a normal pore evolution which results in the formation of high-density Cr,Ca:YAG ceramics without large pores.

The CaO excess, which does not interact with SiO$_2$, segregates at the grain boundaries as has been reported in our previous work, see ref. [21]. The presence of extra CaO at the grain boundaries suppresses grain growth, explaining the absence of abnormal grains in the Ca=0.8 and Ca=1.2 samples. The extra CaO is also responsible for the formation of the Ca-rich phases on the ceramic surface after air annealing (see fig. 3d) due to the high partial pressure of CaO vapor, see Fig. 5. It should be noted that there is evidence that air annealing decreases CaO content, see ref. [21], but, as far as we know, there are not yet data in the literature about the decrease in the calcium-silica phases emerging on the grain boundaries and whether they remain after air annealing. Light scattering on these phases can worsen the optical quality of ceramics.

### 4.2 Influence of Ca$^{2+}$ and Si$^{4+}$ interactions

The SiO$_2$ additive does not only influence the processes under the heating stages, but also the final isothermal stage of YAG ceramics sintering thanks to the ability to generate cation vacancies in the YAG lattice. Kuklja *et al.* proposed that the incorporation of Si$^{4+}$ in the YAG crystal lattice introduces an extra positive charge [27]. To compensate for this charge, cation vacancies are generated. The appearance of cation vacancies increases Y$^{3+}$ and Al$^{3+}$ ions diffusivity in the YAG, which causes an increase in ceramics densification compared to Si-undoped YAG [28].

The incorporation of $Ca^{2+}$ ions in the YAG crystal lattice leads to the formation of $[Ca^{2+}...Si^{4+}]$ neutral complexes, limiting the formation of cation vacancies and thus decreasing ceramics densification. The solubility of $Ca^{2+}$ and $Si^{4+}$ ions in the YAG is very low when they solute individually. $Ca^{2+}$ solubility in the YAG is 300- 400 ppm [29], the solubility limit of $Si^{4+}$ in YAG is up to 600 ppm [30]. However, simultaneous doping of YAG with $Ca^{2+}$ and $Si^{4+}$ significantly increases the solubility of both ions [31], evidenced by the presence of $[Ca^{2+}...Si^{4+}]$ complexes. We assume that $Si^{4+}$ mostly combines with $Ca^{2+}$ when there is enough $Ca^{2+}$ and, in this case, the generation of cation vacancies does not occur. The lack of cation vacancies causes the deterioration of the ceramic densification, which results in a worsening of its optical quality.

The use of $Cr^{4+}$:YAG ceramics as Q-switches requires not only high optical quality, but also high $Cr^{4+}$ ions absorption at laser wavelength (~1 µm). Therefore, the $Cr^{3+}$ to $Cr^{4+}$ transformation was investigated. As shown in Table 3, less than 1% of $Ca^{2+}$ and 2.6% to 6.6% of $Cr^{3+}$ ions participate in the $Cr^{3+} \rightarrow Cr^{4+}$ ions valence transformation. The absorbance of $Cr_D^{4+}$ ions at $\lambda=1030$ nm was close to 1 cm$^{-1}$, which is less than in the case of Cr,Ca:YAG ceramics without $SiO_2$ additives [18,23].

The $Si^{4+}$ ion influence on $Cr^{4+}$ ions concentration is mediated by the formation of $[Ca^{2+}...Si^{4+}]$ neutral complexes, which take away part of $Ca^{2+}$ ions necessary to compensate for the $Cr^{4+}$ positive charge [32]. The dissolution of $Ca^{2+}$ ions in the YAG lattice normally generates neutral complexes with oxygen vacancies. During air annealing, these complexes are destroyed and $Ca^{2+}$ ions recharge $Cr^{3+}$ to $Cr^{4+}$. But the presence of $Si^{4+}$ leads to the formation of neutral complexes $[Ca^{2+}...Si^{4+}]$ in place of $Ca^{2+}$ ions stabilized by oxygen vacancies.

In summary, we propose and discussed two mechanisms of CaO and $SiO_2$ influence on the formation of Ca,Cr:YAG ceramics. The first one consists in the appearance of a liquid phase due to CaO-$SiO_2$ interaction with a specific ratio under the heating stage of the ceramics sintering. Under this way, the abnormal growth of ceramics grains occurs. The second mechanism consists of the mutual consumption of $Si^{4+}$ ions and $Ca^{2+}$ ions under the isothermal stage of ceramic sintering, leading to a decrease of $Cr^{4+}$ concentration. Both mechanisms have a negative effect on the optical and laser properties of the ceramics.

However, some remarks are worth noting: our hypotheses were supported only by the interaction between CaO and $SiO_2$ during sintering, and we did not perform a study of the phase composition and morphology of the samples during the heating stage. Future work will include an evaluation of phase composition and morphology evolution of the samples during the heating stage.

**Conclusions**

The influence of $SiO_2$ and CaO sintering aids on the optical properties and the microstructure of Cr,Ca:YAG ceramics was evaluated in this work. We found that the interactions between CaO and $SiO_2$ compounds, as well as the presence of $Ca^{2+}$ and $Si^{4+}$ ions have a negative impact on the properties of Cr,Ca:YAG ceramics. Decreasing the concentration of CaO led to a decrease in Cr,Ca:YAG ceramics transparency from 81 to 0 % at 1064 nm, moreover, the sample with the lowest concentration of CaO (0.5 at.%) had abnormally large grains and high porosity. The difference in microstructure of Cr,Ca:YAG ceramics was explained through the interaction of CaO and $SiO_2$ on the YAG grain boundaries during the heating stage. The appearance of a liquid phase at 1450°C resulting from the interaction between CaO and $SiO_2$ leads to early collapse of the pore channels that results in the formation of large pores and abnormal grains in the ceramics. The $Ca^{2+}$ - $Si^{4+}$ interaction in the YAG lattice reduces both ceramics density and $Cr^{4+}$ concentrations. $Ca^{2+}$ and $Si^{4+}$ form neutral complexes that inhibit the generation of cation vacancies. The lack in cation vacancies results in a decrease of $Al^{3+}$ and $Y^{3+}$ diffusion during ceramics sintering and to the deterioration of the ceramic's optical quality. The combination of $Ca^{2+}$ and $Si^{4+}$ ions is responsible for the decrease in the concentration of $Cr^{4+}$ ions formed by $Cr^{3+}$ oxidation due to a lack of charge compensator for the extra charge of $Cr^{4+}$. Therefore, the sintering aid $SiO_2$, which is widely used in ceramic technology, should not be used to prepare $Cr^{4+}$:YAG ceramics when great optical quality is desired.

**Acknowledgments:** The authors are grateful to Dr. A.G. Doroshenko and Dr. S.V. Parkhomenko for their help in sintering of the sample. Authors thank Dr. P.V. Mateychenko for providing SEM and EDS measurements at the center for the sharing of scientific equipment "Microscopic and spectroscopic methods for studying the surface of solids" of the STC "Institute for Single Crystals" National Academy of Sciences of Ukraine.

**References**

[1] Angert, N. B., Borodin, N. I., Garmash, V. M., Zhitnyuk, V. A., Okhrimchuk, A. G.,

Siyuchenko, O. G., & Shestakov, A. V. (1988). Lasing due to impurity color centers in yttrium aluminum garnet crystals at wavelengths in the range 1.35–1.45 μm. Soviet Journal of Quantum Electronics, 18(1), 73.

[2] Ubizskii, S., Buryy, O., Börger, A., & Becker, K. D. (2009). Investigation of the chromium ions recharging kinetics in Cr, Mg: YAG crystal during high-temperature annealing. physica status solidi (a), 206(3), 550-561.

[3] Wajler, A., Kozłowska, A., Nakielska, M., Leśniewska-Matys, K., Sidorowicz, A., Podniesiński, D., & Putyra, P. (2014). Nonlinear Absorption of Submicrometer Grain-Size Cobalt-Doped Magnesium Aluminate Transparent Ceramics. Journal of the American Ceramic Society, 97(6), 1692-1695.

[4] Balashov, V. V., Bezotosnyi, V. V., Cheshev, E. A., Gordeev, V. P., Kanaev, A. Y., Kopylov, Y. L., ... & Tupitsyn, I. M. (2019). Composite Ceramic $Nd^{3+}$:YAG/$Cr^{4+}$:YAG Laser Elements. Journal of Russian Laser Research, 40(3), 237-242.

[5] Chaika, M., Vovk, O., Mancardi, G., Tomala, R., & Strek, W. (2020). Dynamics of Yb2+ to Yb3+ ion valence transformations in Yb: YAG ceramics used for high-power lasers. Optical Materials, 101, 109774.

[6]. Tsiumra, V., Krasnikov, A., Zazubovich, S., Zhydachevskyy, Y., Vasylechko, L., Baran, M., ... & Suchocki, A. (2019). Crystal structure and luminescence studies of microcrystalline GGG:$Bi^{3+}$ and GGG:$Bi^{3+}$, $Eu^{3+}$ as a UV-to-VIS converting phosphor for white LEDs. Journal of Luminescence, 213, 278-289.

[7] Chaika, M. A., Mancardi, G., Tomala, R., Stek, W., & Vovk, O. M. (2020). Effects of divalent dopants on the microstructure and conversion efficiency of $Cr^{4+}$ ions in Cr,Me:YAG (Me-Ca, Mg, Ca/Mg) transparent ceramics. Processing and Application of Ceramics, 14(1), 83-89.

[8] Buryy, O. A., Ubiszkii, S. B., Melnyk, S. S., & Matkovskii, A. O. (2004). The Q-switched Nd: YAG and Yb: YAG microchip lasers optimization and comparative analysis. Applied Physics B, 78(3-4), 291-297.

[9] Feldman, R., Shimony, Y., & Burshtein, Z. (2003). Dynamics of chromium ion valence transformations in Cr,Ca:YAG crystals used as laser gain and passive Q-switching media. Optical

Materials, 24(1-2), 333-344.

[10] Ikesue, A., Yoshida, K., Yamamoto, T., & Yamaga, I. (1997). Optical scattering centers in polycrystalline Nd: YAG laser. Journal of the American Ceramic Society, 80(6), 1517-1522.

[11] Chretien, L., Bonnet, L., Boulesteix, R., Maitre, A., Salle, C., & Brenier, A. (2016). Influence of hot isostatic pressing on sintering trajectory and optical properties of transparent Nd: YAG ceramics. Journal of the European Ceramic Society, 36(8), 2035-2042.

[12] Liu, B., Li, J., Ivanov, M., Liu, W., Ge, L., Xie, T., ... & Guo, J. (2015). Diffusion-controlled solid-state reactive sintering of Nd:YAG transparent ceramics. Ceramics International, 41(9), 11293-11300.

[13] Vorona, I. O., Yavetskiy, R. P., Shpilinskaya, O. L., Kos'yanov, D. Y., Doroshenko, A. G., Parkhomenko, S. V., ... & Tolmachev, A. V. (2015). The effect of residual porosity on the optical properties of $Y_3Al_5O_{12}$:$Nd^{3+}$ laser ceramics. Technical Physics Letters, 41(5), 496-499..

[14] Chaika, M. A., Tomala, R., Strek, W., Hreniak, D., Dluzewski, P., Morawiec, K., ... & Lesniewska-Matys, K. (2019). Kinetics of $Cr^{3+}$ to $Cr^{4+}$ ion valence transformations and intra-lattice cation exchange of $Cr^{4+}$ in Cr,Ca:YAG ceramics used as laser gain and passive Q-switching media. The Journal of Chemical Physics, 151(13), 134708.

[15] Fabrichnaya, O., Seifert, H. J., Weiland, R., & Ludwig, T. (2001). Phase Equilibria and Thermodynamics in the YO-AlO–SiO System.

[16] Stevenson, A. J., Li, X., Martinez, M. A., Anderson, J. M., Suchy, D. L., Kupp, E. R., ... & Messing, G. L. (2011). Effect of $SiO_2$ on densification and microstructure development in Nd:YAG transparent ceramics. Journal of the American Ceramic Society, 94(5), 1380-1387.

[17] M. A. Chaika, O. M. Vovk, N. A. Safronova, A. G. Doroshenko, S. V. Parkhomenko, A. V. Tolmachev, (2016) Mutual influence of additives of Ca and Si on properties of Cr-doped YAG ceramics, Functional materials 23(3), 398–403.

[18] Zhou, T., Zhang, L., Yang, H., Qiao, X., Liu, P., Tang, D., & Zhang, J. (2015). Effects of sintering aids on the transparency and conversion efficiency of $Cr^{4+}$ ions in Cr:YAG transparent ceramics. Journal of the American Ceramic Society, 98(8), 2459-2464.

[19] O. V.Gayduk, (2017) Control of chromium dopant content in optical ceramics Cr:YAG, Functional materials 24(2), 318–321.

[20] L. Gurvich, I. Veits, C. Alcock, Thermodynamics properties of individual substances, New York, NY (US); Hemisphere Publishing Corp., 1989.

[21] Chaika, M. A., Dluzewski, P., Morawiec, K., Szczepanska, A., Jablonska, K., Mancardi, G., Doroshenko, A. G. (2019). The role of $Ca^{2+}$ ions in the formation of high optical quality $Cr^{4+}$,Ca:YAG ceramics. Journal of the European Ceramic Society, 39(11), 3344-3352.

[22] Balabanov, S. S., Bykov, Y. V., Egorov, S. V., Eremeev, A. G., Gavrishchuk, E. M., Khazanov, E. A., Zelenogorskii, V. V. (2013). Yb:$(YLa)_2O_3$ laser ceramics produced by microwave sintering. Quantum Electronics, 43(4), 396.

[23] Chaika, M. A., Dulina, N. A., Doroshenko, A. G., Parkhomenko, S. V., Gayduk, O. V., Tomala, R., ... & Vovk, O. M. (2018). Influence of calcium concentration on formation of tetravalent chromium doped $Y_3Al_5O_{12}$ ceramics. Ceramics International, 44(12), 13513-13519.

[24] Chaika, M., Paszkowicz, W., Strek, W., Hreniak, D., Tomala, R., Safronova, N., ... & Vovk, O. (2019). Influence of Cr doping on the phase composition of Cr, Ca: YAG ceramics by solid state reaction sintering. Journal of the American Ceramic Society, 102(4), 2104-2115.

[25] A. Ragulya, V. Skorokhod, Consolidated nanostructured materials, Naukova Dumka, Kiev 374.

[26] Rankin, G. A. (1915). The ternary system $CaO$-$Al_2O_3$-$SiO_2$, with optical study by FE Wright. American Journal of Science, (229), 1-79.

[27] Kuklja, M. M. (2000). Defects in yttrium aluminium perovskite and garnet crystals: atomistic study. Journal of Physics: Condensed Matter, 12(13), 2953.

[28] Kochawattana, S., Stevenson, A., Lee, S. H., Ramirez, M., Gopalan, V., Dumm, J., ... & Messing, G. L. (2008). Sintering and grain growth in $SiO_2$ doped Nd: YAG. Journal of the European Ceramic Society, 28(7), 1527-1534.

[29] Schuh, L., Metselaar, R., & de With, G. (1989). Electrical transport and defect properties of Ca-and Mg-doped yttrium aluminum garnet ceramics. Journal of applied Physics, 66(6), 2627-2632.

[30] Y Wang, Y., Zhang, L., Fan, Y., Luo, J., McCready, D. E., Wang, C., & An, L. (2005). Synthesis, characterization, and optical properties of pristine and doped yttrium aluminum garnet nanopowders. Journal of the American Ceramic Society, 88(2), 284-286.

[31] Kuru, Y., Onur Savasir, E., Zeynep Nergiz, S., Oncel, C., & Ali Gulgun, M. (2008). Enhanced co-solubilities of Ca and Si in YAG ($Y_3Al_5O_{12}$). physica status solidi c, 5(10), 3383-3386..

[32] Chaika, M. A., Vovk, O. M., Doroshenko, A. G., Klochkov, V. K., Mateychenko, P. V., Parkhomenko, S. V., & Fedorov, O. G. (2017). Influence of Ca and Mg doping on the microstructure and optical properties of YAG ceramics. Functional Materials. 24(2), 237-243.